# Influence of the martensitic transformation kinetics on the magnetocaloric effect in Ni-Mn-In

L. Pfeuffer[1], T. Gottschall[2], T. Faske[1], A. Taubel[1], F. Scheibel[1], A. Y. Karpenkov[1,3], S. Ener[1], K. P. Skokov[1] and O. Gutfleisch[1]

[1] *Technical University of Darmstadt, Institute of Materials Science, 64287 Darmstadt, Germany*

[2] *Dresden High Magnetic Field Laboratory (HLD-EMFL), Helmholtz-Zentrum Dresden-Rossendorf, 01328 Dresden, Germany*

[3] *National Research South Ural State University, 454080 Chelyabinsk, Russian Federation*

**Abstract**

The inverse magnetocaloric effect (MCE) in Ni-Mn-based Heusler compounds occurs during the magnetostructural transition between low-temperature, low-magnetization martensite and high-temperature, high-magnetization austenite. In this study, we analyze the metamagnetic transformation of a $Ni_{49.8}Mn_{35}In_{15.2}$ compound by simultaneous adiabatic temperature change $\Delta T_{ad}$ and strain $\Delta l/l_0$ measurements in pulsed magnetic fields up to 10 T. We observe a $\Delta T_{ad}$ of -10 K and a $\Delta l/l_0$ of -0.22 % when the reverse martensitic transition is fully induced at a starting temperature of 285 K. By a variation of the magnetic field-sweep rates between 316 $Ts^{-1}$, 865 $Ts^{-1}$ and 1850 $Ts^{-1}$, the transitional dynamics of the reverse martensitic transformation have been investigated. Our experiments reveal an apparent delay upon the end of the reverse martensitic transformation at field rates exceeding 865 $Ts^{-1}$ which is related to the annihilation of retained martensite. As a consequence, the field hysteresis increases and higher fields are required to saturate the transition. In contrast, no time-dependent effects on the onset of the reverse martensitic transformation were observed in the studied field-sweep range. Our results demonstrate that kinetic effects in Heusler compounds strongly affect the magnetic cooling cycle, especially when utilising a multicaloric "exploiting-hysteresis cycle" where high magnetic field-sweep rates are employed.

1. Introduction

Since the discovery of the giant magnetocaloric effect (MCE) of $Gd_5(Si_xGe_{1-x})_4$ in 1997 [1], other first-order materials such as $La(Fe,Si)_{13}$ [2–4], $Fe_2P$-type [5, 6] and Heusler compounds [7–10] have found great attention in the magnetic refrigeration community. However, the intrinsic thermal hysteresis accompanying the magnetostructural transformation significantly limits the utilization of the materials isothermal entropy change $\Delta s_T$ and adiabatic temperature change $\Delta T_{ad}$ in cyclic operation [11, 12]. For multicaloric materials, the addition of mechanical fields such as hydrostatic pressure [13] or uniaxial load [14] upon demagnetization is one approach to minimize the effective thermal hysteresis. We recently proposed an alternative concept to increase reversibility, which is based on the exploitation of thermal hysteresis rather than avoiding it [15]. In this case, the material is trapped in its ferromagnetic state after applying and removing the magnetic field by the accepted hysteresis and is sequentially transformed back to its initial non-magnetic state by uniaxial stress.

Accordingly, potential candidate materials for the so-called "exploiting-hysteresis cycle" require a high susceptibility to magnetic field and uniaxial load. Thereby, the first-order transition has to exhibit an inverse magnetocaloric and a conventional elastocaloric effect or the vice versa combination to ensure cyclability [16]. For Ni-Mn-X-(Co) based Heusler compounds with X = In [13, 17, 18], Sb [19, 20] or Sn [21–23], large inverse magnetocaloric and conventional elastocaloric effects based on their martensitic transformation have been reported. While the magnetic field stabilizes the high-temperature, high-magnetization austenite, the uniaxial load results in a preferential formation of low-temperature, low-magnetization martensite, because of its lower crystal symmetry [24]. In addition, a sufficiently large thermal hysteresis to avoid demagnetization upon magnetic field removal is present in a variety of Ni-Mn-X-(Co) metamagnetic shape-memory alloys [25].

The suppressed demagnetization when the magnetic field is removed has the major advantage that the magnetic field does not have to be maintained during heat transfer. Accordingly, the large volume of expensive permanent magnets, which are typically used for magnetic refrigeration can be significantly reduced and higher, as more focused, magnetic fields can be reached [26, 27]. In addition, the decoupling of the (de)magnetization and heat transfer process allows the utilization of shorter exposure times of the magnetocaloric material to the magnetic field, which is limited in conventional active magnetic regeneration by the operational frequency of usually 1 Hz [28]. As a consequence, a detailed understanding of the magnetocaloric materials response to short-time magnetic field exposure and with this high magnetic field-sweep rates is inevitable.

For this purpose, we have studied the magnetostructural response of the promising metamagnetic shape-memory alloy Ni-Mn-In to varying high magnetic field-sweep rates, generated in a solenoid, by simultaneous $\Delta T_{ad}$ and strain $\Delta l/l_0$ measurements. The combination of both signals enabled us to circumvent the frequently discussed problem of thermocouple response time in first-order magnetocaloric materials, and allowed the analysis of the transitional dynamics in addition to the direct determination of the MCE [29–32]. With this approach, a new detailed insight into the reverse martensitic transformation kinetics in metamagnetic Ni-Mn-In was obtained.

2. **Experimental details**

Polycrystalline, nominally composed $Ni_{49.8}Mn_{35}In_{15.2}$ was synthesized by manifold arc melting of high purity Ni (99,97 %), Mn (99,99 %) and In (99,99 %). The chemically homogenized ingot was subsequently annealed in a quartz tube under Ar atmosphere at 900 °C for 24 h, followed by rapid quenching in water.

The simultaneous $\Delta l/l_0$ and $\Delta T_{ad}$ measurements were performed in pulsed magnetic fields of 2, 5 and 10 T in a solenoid magnet at the Dresden High Magnetic Field Laboratory (HLD). The maximum fields of 2, 5 and 10 T were always reached after 13 ms, with maximum magnetic-field rates of 316, 865 and 1850 Ts$^{-1}$, respectively. For the detection of the adiabatic temperature change $\Delta T_{ad}$, a differential T-type thermocouple with a single wire thickness of 25 μm was fixed between two parts of the sample by a thermally conductive epoxy. The strain determination was carried out by a linear pattern strain gage of 0.79 mm gage length and 1.57 mm grid width glued on the sample surface. The electrical resistance of the strain gage was determined via a function generator and a digital lock-in technique. The gage direction was positioned perpendicular to the applied magnetic field, which was recorded by a pick-up coil. For the determination of the absolute temperature of the sample, a Pt100 temperature sensor was used. A more detailed description of the setup can be found in [30] and [31].

All simultaneous measurements of $\Delta T_{ad}$ and $\Delta l/l_0$ were executed after 25 training pulses to ensure a repeatable response of the sample upon the field-induced transformation. In addition, a discontinuous temperature protocol was used in order to bring the material back to a defined initial state after application of the magnetic field [29, 33]. For this purpose, the sample was heated up to 310 K (fully austenitic state) and cooled down to 240 K (fully martensitic state) before setting the sample temperature for the measurement [34].

Isofield curves of magnetization and strain of the prepared sample were carried out in VSM mode of a Quantum Design PPMS-14 T using a heating and cooling rate of 2 Kmin$^{-1}$. In accordance with pulsed-field measurements, the gage direction was positioned perpendicular to the magnetic field [16].

Temperature-dependent optical microscopy was executed at a Zeiss Axio Imager.D2M equipped with a LN$_2$ cryostat. A heating and cooling rate of 2 Kmin$^{-1}$ was used.

Temperature-dependent X-ray measurements were collected on a custom-built diffractometer in transmission geometry (Mo K$_\alpha$ radiation, $\lambda_1 = 0.070932$ nm, $\lambda_2 = 0.071332$ nm, MYTHEN2 R 1K detector (Dectris Ltd.), 2θ range from 7 to 57°, step size of 0.009°). A detailed description of the diffractometer can be found in [35]. A small piece of the sample was crushed into powder with particle sizes < 40 µm. To ensure the release of the deformation induced stresses, the powder was annealed in a fused silica tube under Ar atmosphere for 7 days at 850 °C and subsequently quenched in water. The annealed powder was mixed with a NIST640d standard reference silicon powder for correction of geometric errors and glued onto a graphite sheet. The temperature was controlled by means of a closed-cycle helium cryofurnace (SHI Cryogenics Group) in the range from 310 to 240 K.

## 3. Results

### A. Structural Properties

Figure 1 shows the XRD patterns of the synthesized Ni$_{49.8}$Mn$_{35}$In$_{15.2}$ sample at 310 K and 240 K. The diffractogram at 310 K reveals a solely austenitic phase having a L2$_1$ structure with a lattice constant of a = 0.60027 nm. At 240 K the sample is almost entirely in a 3 M modulated martensite state with a monoclinic unit cell having the lattice parameters a = 0.43947 nm, b = 0.56330 nm, c = 3*0.43307 nm and β = 92.691°. The change of the lattice parameters indicates a highly anisotropic structural distortion upon the martensitic transformation, resulting in an overall volume decrease of 0.97 %. The lattice parameters and the corresponding volume change are in good agreement with the data reported in [36, 37].

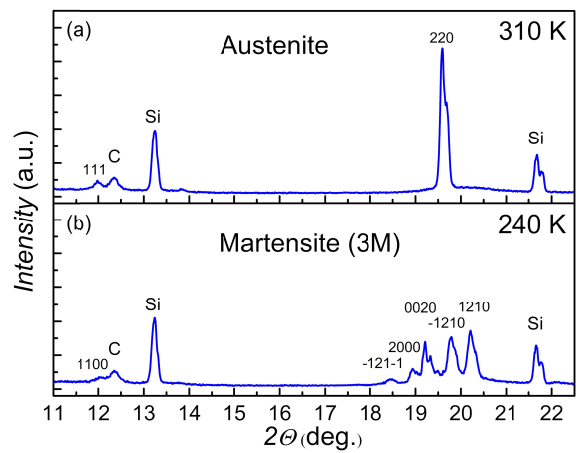

Figure 1: XRD patterns of Ni$_{49.8}$Mn$_{35}$In$_{15.2}$ powder at (a) 310 K (L2$_1$ austenite) and (b) 240 K (3 M modulated martensite). Austenite reflections are labelled with their hkl indices. For the martensite reflections, the respective hklm indices according to (3+1)D superspace symmetry are *used. The Si reflections result from the NIST640d reference and the C reflection from the graphite sheet utilized for gluing the powder.*

## B. Isofield Measurements

Figure 2 summarizes the simultaneous magnetization and strain measurements for heating and cooling in magnetic fields of 0.05 T up to 10 T. A magnetostructural transition from high-temperature, high-magnetization austenite to low-temperature, low-magnetization martensite can be observed at 281 K in 0.05 T (see Figure 2(a)). The corresponding reverse transformation occurs at 288 K resulting in a thermal hysteresis of 7 K. An increase of the magnetic field stabilizes the high-magnetization austenite phase leading to a shift of the austenite-to-martensite transition temperature by -4.87 KT$^{-1}$, whereas the martensite-to-austenite transition temperature changes by -4.27 KT$^{-1}$. As a consequence, the thermal hysteresis expands to 13 K in a magnetic field of 10 T. It is worth mentioning, that the total entropy change $\Delta s_t$ is lowered, when the martensitic transformation is shifted towards lower temperatures by magnetic field or chemical composition [38]. This effect is caused by a rising magnetic entropy change $\Delta s_{mag}$, which results from the increasing magnetic entropy contribution of the austenite at lower temperatures and counteracts the structural part $\Delta s_{lat}$ of $\Delta s_t$ according equation (1).

$$\Delta s_t = |\Delta s_{lat}| - |\Delta s_{mag}| \quad (1)$$

In equation (1), the electronic entropy contribution $\Delta s_{el}$ is not considered, as *Kihara et al.* [39] showed that the influence on $\Delta s_t$ is negligibly small for Heusler compounds. Besides the inverse MCE at the magnetostructural transition, a conventional MCE is present at the Curie temperature of the austenite $T_C^A$ = 312 K.

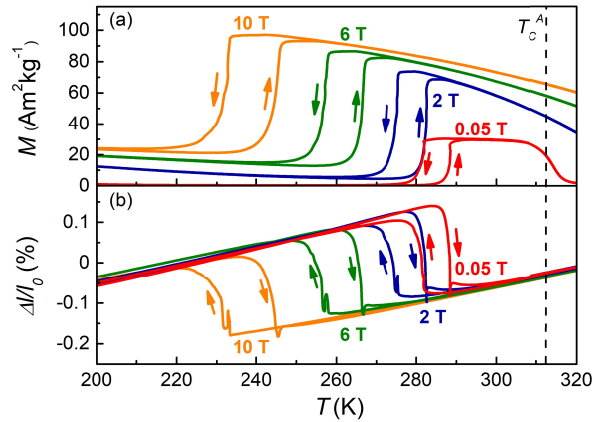

*Figure 2: Isofield measurements of the magnetization M (a) and the strain $\Delta l/l_0$ (b) in a magnetic field of 0.05 T, 2 T, 6 T and 10 T.*

The difference between the first- and second-order transition becomes apparent in Figure 2(b). While no structural change takes place at $T_C^A$, the strong coupling of magnetic moments and crystal lattice results in a macroscopic length change $|\Delta l/l_0|$ of 0.22 % upon the martensitic transformation at 0.05 T. This transition strain value is comparable or even higher than reported in [36, 37, 40–42] for polycrystalline Ni-Mn-In-(Co) compounds. The length expansion upon the growth of the low-volume martensite phase can be explained by the anisotropic change of the lattice parameters (see Figure 1). With increasing magnetic field, the transition strain $|\Delta l/l_0|$ is reduced to 0.18 % in 10 T. The decrease in $|\Delta l/l_0|$ can be ascribed to the decline of $\Delta s_t$ at lower transformation temperatures, which will be discussed later in more detail.

*Pathak et al.* [43] showed for Ni-Mn-In-(Si) that also microstructural features such as grain size and orientation with respect to the gage and magnetic field direction can significantly influence the transition-strain behavior. In the present study, the underlying microstructure and its evolution were investigated by temperature-dependent optical microscopy (see Figure 3). At 300 K, in the full austenite state, large columnar grains of 100-700 µm diameter can be observed, which are characteristic for arc-molten Ni-Mn-In-(Co) [41]. Note that all strain measurements were carried out perpendicular to the solidification direction and to the magnetic field as shown in the schematic drawing in Figure 3. Upon cooling, the martensitic transformation can be noticed by the formation of a distinct surface relief. Thereby, the martensite nucleates and grows preferably along grain boundaries and in 45° angles within the grains, which is indicated by the micrograph in the mixed-phase state at 283 K. The resulting variants in the solely martensitic state are displayed in the micrograph at 270 K.

A further influence on the strain upon the martensitic transformation can result from the magnetocrystalline anisotropy of the martensite below its Curie temperature $T_C^M$ [44]. Also related to the magnetostructural transition, the strain signal in Figure 2(b) shows the occurrence of humps upon the martensite formation and the austenite finishing process. Similar anomalies in $\Delta l/l_0$ were already reported for Ni-Mn-In [40] and Ni-Mn-Ga [45] and might result from microstructural changes during the nucleation and annihilation of martensite, elastic softening, active defects or microscopic features such as microcracks [46–48].

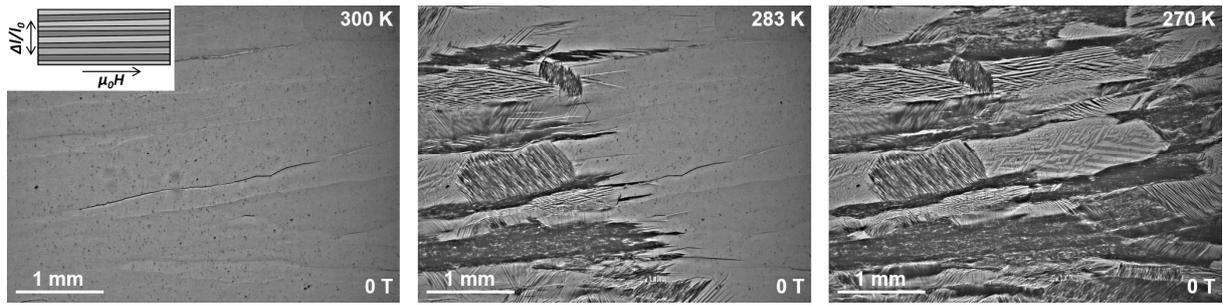

*Figure 3: Temperature-dependent optical microscopy upon cooling in zerofield. Images were taken at 300 K (full austenite), 283 K (mixed state) and 270 K (full martensite). The schematic in the top left corner visualizes the gage and field direction with respect to the grain orientation.*

### C. Pulsed-Field Measurements

Figure 4(a) illustrates the temporal evolution of $\Delta T_{ad}$ and $\Delta l/l_0$ in a magnetic-field pulse of 10 T when the sample has an initial temperature $T_{start}$ of 285 K. Thereby, the maximum magnetic field is reached after 13 ms with a maximum field-sweep rate of 1850 Ts$^{-1}$. In contrast to that, the field removal is significantly slower. The field dependence of $\Delta T_{ad}$ and $\Delta l/l_0$ for $T_{start}$ = 285 K is shown in Figure 4(b). At this temperature, the sample was still martensitic in zerofield, but close to the austenite formation, as $T_{start}$ was always approached from 240 K (full martensite). With the application of the magnetic field, the sample transforms into austenite as soon as the austenite starting field $H_{As}$ is reached. Figure 4(b) implies that $H_{As}$ differs significantly for $\Delta T_{ad}$ compared to $\Delta l/l_0$. This effect can be ascribed to the different scale of the $\Delta T_{ad}$- and $\Delta l/l_0$-axis and a delay of the thermocouple, which will be discussed later in more detail. At a starting temperature of 285 K, the sample shows a $\Delta T_{ad}$ of -10 K and a $\Delta l/l_0$ of -0.22 % when the magnetostructural transformation is completed. The detected strain is in excellent agreement with the isofield measurements, though different stimuli act as driving forces for the transition. Figure 4(b) shows that the transformation to the austenite state is finished at a magnetic field $H_{Af}$ of approximately 7 T. The increase of the temperature above this completion field relates to the conventional MCE of the ferromagnetic austenite phase, which has been created during the pulse. The same effect leads to the linear decrease of the relative temperature to -11.5 K down to 2.2 T when the magnetic field is removed. This corresponds to a further increase of $|\Delta T_{ad}|$ by 15 %. Since this effect is just related to the magnetization, a constant $\Delta l/l_0$ is observed until the martensitic transformation begins below 2.2 T. Thereby, the transition back to the martensite state is partially hindered because of the thermal hysteresis, resulting in a residual $\Delta T_{ad}$ of -3.4 K and $\Delta l/l_0$ of -0.057 % after the field is completely taken off.

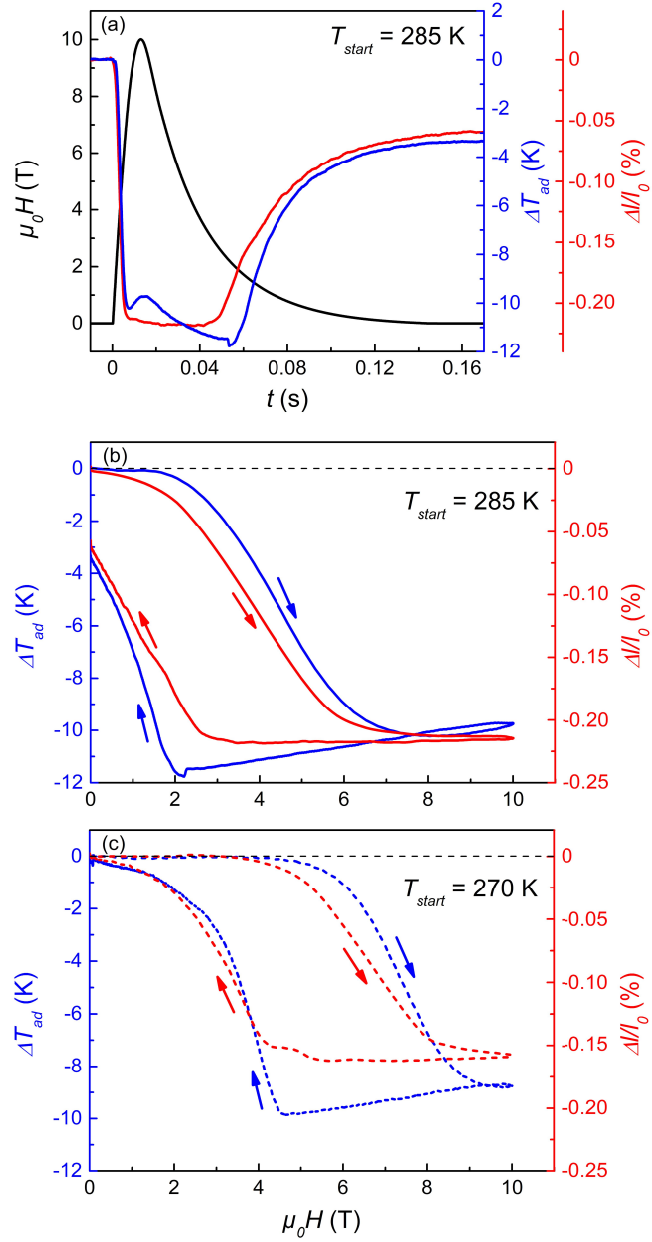

*Figure 4: (a) Adiabatic temperature change $\Delta T_{ad}$ (blue) and strain $\Delta l/l_0$ (red) as a function of time for a magnetic-field pulse of 10 T and an initial sample temperature $T_{start}$ = 285 K. The black line represents the time dependence of the magnetic field. (b) Corresponding magnetic field-dependent evolution of $\Delta T_{ad}$ and strain $\Delta l/l_0$ at $T_{start}$ = 285 K. (c) Magnetic field-dependent evolution of $\Delta T_{ad}$ and strain $\Delta l/l_0$ at an initial sample temperature of $T_{start}$ = 270 K.*

Figure 4(c) shows that the reversibility can be increased by lowering the initial sample temperature. A full recovery of the martensite can be obtained at $T_{start}$ = 270 K, which is visible for both, $\Delta T_{ad}$ and $\Delta l/l_0$. However, higher magnetic fields are needed to induce and complete the transformation [49].

Thereby, $|\Delta T_{ad}|$ and $|\Delta l/l_0|$ exhibit significantly lower values than at a $T_{start}$ of 285 K for the fully induced martensite-to-austenite transformation. This effect can be ascribed to the increasing magnetic entropy change $\Delta s_{mag}$ when $T_{start}$ is reduced. Consequently, the total entropy change $\Delta s_t$ decreases (see equation (1)), which corresponds to a decay of the maximum achievable adiabatic temperature change $|\Delta T_{ad}|$ according

$$\Delta T_{ad} \simeq -\frac{T}{c_p} \Delta s_t \qquad (2)$$

with the background heat capacity $c_p$ [38, 50]. Equation (2) shows that also the absolute temperature $T$ of the sample contributes to the decline of $|\Delta T_{ad}|$ at lower temperatures, while the temperature dependence of $c_p$ for Ni-Mn-In causes only a weak opposing behavior as the Dulong-Petit-limit is reached for the studied temperature range [51]. The decay of $|\Delta l/l_0|$ upon lowering $T_{start}$ can also be related to the decrease of $\Delta s_t$ according to the Clausius-Clapeyron equation

$$\Delta s_t = -\left(\frac{1}{\rho}\right) * \left(\frac{\Delta l}{l_0}\right) * \left(\frac{dT_t}{d\sigma}\right)^{-1} \qquad (3)$$

with the mass density $\rho$ and the shift of the transition temperature with applied stress $(dT_t/d\sigma)$. Note that the latter is temperature-dependent and therefore a further potential influencing variable for $|\Delta l/l_0|$. The decay of $|\Delta l/l_0|$ at lower temperatures is in good agreement with the isofield curves of strain, shown in Figure 2(b). After careful consideration of the isofield measurements (see Figure 2(b)), small anomalies in the strain signal can be noticed upon the back transformation to the martensite. Thus, the occurrence of a shoulder in $\Delta l/l_0$ can be seen at the onset of the back transformation for $T_{start}$ = 270 K. At $T_{start}$ = 285 K, the anomaly occurs in a way, that the martensite start field $H_{Ms}$ of the $\Delta l/l_0$ signal, at 2.6 T, is approximately 0.7 T higher than the one of the corresponding $\Delta T_{ad}$ signal. In general, a response of the strain gage can be noticed in all pulsed-field measurements ahead of the thermocouple upon the back transformation to the martensite. The origin of this behavior we ascribe to the already discussed anomalies of $\Delta l/l_0$. In addition, it should be considered that the strain is measured at the sample surface, while $\Delta T_{ad}$ represents the behavior of the volume. Further investigations will be done to clarify the origin of this difference. In contrast to the isofield measurements, no anomalies of $\Delta l/l_0$ occur during the formation of the austenite in pulsed fields, which indicates an influence of the stimulus on the transition behavior.

In Figure 5, $\Delta T_{ad}$ and $\Delta l/l_0$ are plotted for different initial temperatures in magnetic field pulses of 2 T, 5 T and 10 T. Thereby, both quantities were determined at the magnetic field in which the transition could be fully induced to cancel out the influence of the conventional MCE on $\Delta T_{ad}$. In case of an incomplete transformation, the values were taken at the maximum magnetic field. The grey area between 287.1 K and 290 K displays the main austenite transformation regime in zerofield. Figure 5 shows that $\Delta T_{ad}$ and $\Delta l/l_0$ have their maxima when the sample is initially in full martensite state, but as close as possible to the austenite start temperature $A_s$. For the measurement in 2 T, this is the case at 287 K. As soon as the initial sample temperature $T_{start}$ is lower, e.g. 283 K, a decrease of $|\Delta T_{ad}|$ and $|\Delta l/l_0|$ is observed. It should be mentioned, that a magnetic field of 2 T is not sufficient to complete the transformation. The same holds true for magnetic fields of 5 T when $T_{start}$ is lower than 285 K (in case of $T_{start}$ = 285 K a full transformation can be achieved, which will be discussed later in more detail). Hence, for both magnetic fields, 2 T and 5 T, the decay of $|\Delta T_{ad}|$ and $|\Delta l/l_0|$ can be related to the decrease of the transformed phase fraction when $T_{start}$ is even further below $A_s$, as well as to a decrease of the total entropy change $\Delta s_t$. In contrast to that, the drop of $|\Delta T_{ad}|$ and $|\Delta l/l_0|$ for the measurement at 289 K in 2 T solely results from the reduction of the transformed phase fraction, as the sample was already partially austenitic before the field pulse. In 10 T, the transformation can be fully induced from 285 K down to 265 K. Hence, the decrease of $|\Delta T_{ad}|$ and $|\Delta l/l_0|$ upon lowering $T_{start}$ from 285 K to 265 K is solely related to the aforementioned decay of $\Delta s_t$ and both parameters show a more plateau-like behavior. For $T_{start}$ being below 265 K, additionally a reduction of the transformed volume fraction contributes to the decrease in $|\Delta T_{ad}|$ and $|\Delta l/l_0|$ as it is the case in 2 T and 5 T. When $T_{start}$ is close to the austenite finish temperature $A_f$, e.g. at 290 K, $\Delta T_{ad}$ is dominated by the conventional MCE and becomes accordingly positive. The slightly negative $\Delta l/l_0$ of -0.003 % in 10 T can be related to a small fraction of rest martensite, which is still transformed at this temperature.

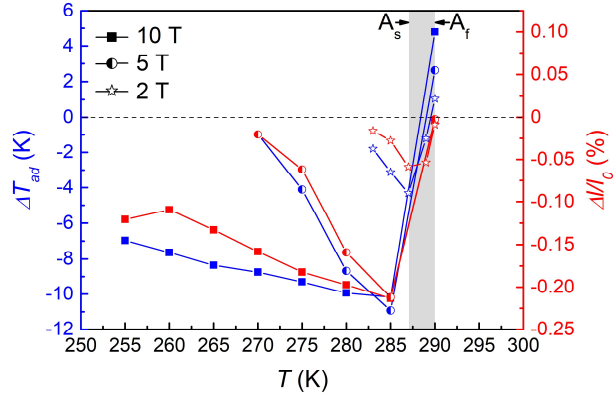

*Figure 5: Maximum adiabatic temperature change $\Delta T_{ad}$ (blue) and strain $\Delta l/l_0$ (red) at magnetic fields of 2 T (stars), 5 T (circles) and 10 T (squares). The grey area indicates the temperature range of the austenite formation in zerofield.*

It is worth pointing out, that the maximum $|\Delta T_{ad}|$ and $|\Delta l/l_0|$ reached at a $T_{start}$ of 285 K, coincides for 5 T and 10 T, which indicates a fully induced transformation for the 5 T pulse. Considering field dependencies of $\Delta T_{ad}$ shown in Figure 4(b), this finding is not intuitive as for the corresponding 10 T pulsed-field measurement about 7 T are needed to complete the transformation. Consequently, a lower $|\Delta T_{ad}|$ and $|\Delta l/l_0|$ would be expected in 5 T than in 10 T.

For the explanation of this behavior, time-dependent effects of the first-order transition should be investigated. In Figure 6(a), the magnetic field pulses are plotted as a function of time. Thereby the maximum magnetic field was always reached after 13 ms with maximum field-sweep rates $\mu_0(dH/dt)$ of 1850 Ts$^{-1}$, 865 Ts$^{-1}$ and 316 Ts$^{-1}$ for 10 T, 5 T and 2 T, respectively (see inset of Figure 6 (a)). In contrast to that, the field removal is considerably slower and the field rates coincide. Accordingly, different magnetic-field pulses at a constant initial starting temperature allow studying kinetic effects of the field-induced reverse martensitic transformation.

Figure 6(b) and (c) illustrate the $\Delta T_{ad}$ and $\Delta l/l_0$ response in magnetic fields of 2 T, 5 T and 10 T at $T_{start} = 285$ K. As discussed earlier, the 10 T pulse with a maximum field rate of 1850 Ts$^{-1}$ displays a complete transformation at ca 7 T. For the 5 T pulse having a maximum field rate of 865 Ts$^{-1}$, a conventional MCE of the austenite and a saturation of the strain signal cannot be observed. However, the signal of both, $\Delta T_{ad}$ and $\Delta l/l_0$, for the maximum field strength of 5 T coincides with the respective signal of the field removal path for the completed transition of the 10 T pulse. This indicates, that for a topmost field rate of 865 Ts$^{-1}$ a complete transformation can be achieved in 5 T. Moreover, the good agreement of the signals in 5 T and 10 T upon field removal proves the accordance of the initial sample temperature for both pulses. The 2 T measurement with a maximum field-sweep rate of 316 Ts$^{-1}$ shows a distinct minor-loop behavior, as in the provided field strength only a fraction of martensite can be transformed.

Comparing the temperature evolution for 2 T, 5 T and 10 T in Figure 6(b), a slight increase of the austenite start field $H_{As}$ can be observed with rising field-sweep rate. For the $\Delta l/l_0$ response in Figure 6(c) such a dependence doesn't occur and the values for $H_{As}$ coincide. The different behavior of both quantities originates from a small increasing delay of the thermocouple when the field-sweep rate is enlarged, which is visible for the $\Delta T_{ad}$ measurements at $T_C^A = 312\ K$ (see inset Figure 6(b)). As the MCE at $T_C^A$ is of second-order, $\Delta T_{ad}$ must be the same upon field application and removal. However, the signal shows a rising hysteresis with higher rates revealing the retardation of the thermocouple. For the strain gage no delay was found (not shown here) for the studied field-sweep rates between 316 Ts$^{-1}$ and 1850 Ts$^{-1}$. Hence, the $\Delta l/l_0$ signal clearly demonstrates, that no kinetic effect on the onset of the reverse martensitic transformation can be noticed for the studied field rates. This finding can be related to recent studies which show that no nucleation of austenite is required for the onset of the reverse martensitic transformation [52, 53].

Moreover, Figure 6(b) and (c) display a substantial influence of field rates higher than 865 Ts$^{-1}$ during the reverse martensitic transformation. Thus, a much bigger slope of $\Delta T_{ad}$ and $\Delta l/l_0$ can be observed during the transformation process in 865 Ts$^{-1}$ compared to 1850 Ts$^{-1}$. This behavior indicates, that at field rates significantly higher than 865 Ts$^{-1}$ the phase transition cannot follow the magnetic field anymore. Accordingly, higher magnetic fields are needed to complete the transformation when the field rate exceeds a critical value, as observed in 1850 Ts$^{-1}$ for the 10 T pulse. As a consequence, an enlargement of the field hysteresis can be observed upon rising sweeping speeds.

Based on the in-situ microscopical investigations in the present and previous studies [54, 55], we assume that the dynamical effects originate from the annihilation of retained martensite in the newly formed austenite matrix. Figure 7 depicts the micrographs of the arc-molten $Ni_{49.8}Mn_{35}In_{15.2}$ upon cooling near $M_s$, in the full martensite state and near $A_f$ during the subsequent heating. A video of the entire transformation upon cooling and heating is provided in the supplementary material. Near $M_s$, the martensite starts to form in the austenite matrix at the grain boundaries and by distinct parallel lines within the grains. On cooling, further nucleation and growth of martensite takes place until the transformation is completed at $M_f$. During the subsequent heating, the reverse process can be observed. The martensite shrinks and fully annihilates at $A_f$. Thereby, the retained martensite in the newly formed austenite matrix near $A_f$ coincides well with the martensite initially formed near $M_s$. This microstructural coincidence suggests that similar to the nucleation of the martensite upon cooling or field removal, the disappearance of the martensite upon heating or field application doesn't occur instantaneously because of local phase incompatibilities. Consistent with our argument, *Xu et al.* [56, 57] observed in Ni-Mn-In-Co that with increasing field rates the martensite start field $H_{Ms}$ decreases upon field removal while the $H_{Af}$ increases upon field application. However, further investigations are needed for a detailed understanding of the underlying mechanisms of the transformation delay at field rates higher than 865 Ts$^{-1}$. Especially, the specific microstructure of the material can play an essential role due to differences in the defect density. It should be noticed, that in the present study the sweeping rates coincide upon field removal and thus no distinct difference in $H_{Ms}$ can be observed in the $\Delta l/l_0$ signal (see Figure 6(c)). The same holds true for the onset of the martensitic transformation in the $\Delta T_{ad}$ signal (see Figure 6(b)), which is not affected by the thermocouple response time due to a considerably slower field rate. Hence, the already discussed difference in $H_{Ms}$ when the $\Delta T_{ad}$ and $\Delta l/l_0$ signal are compared (see Figure 4(b)) is not related to a delay of the thermocouple.

We extend here now on our earlier investigations [29] on the time-dependent transformation of Ni-Mn-In. *Gottschall et al.* [29] showed by $\Delta T_{ad}$ measurements that the onset is the dominating factor for kinetic effects of the reverse martensitic transformation, when a 5 T pulse with a field-sweep rate of 750 Ts$^{-1}$ is compared to the slow field rates in a Halbach setup (0.7 Ts$^{-1}$) and a superconducting magnet (0.0125 Ts$^{-1}$). In this study, we show by simultaneous $\Delta T_{ad}$ and $\Delta l/l_0$ measurements, that in pulsed fields having field rates between 316 Ts$^{-1}$ and 1850 Ts$^{-1}$ no change of the onset occurs. Instead, we could demonstrate that dynamic effects arise upon the annihilation of retained martensite when a field-sweep rate of 865 Ts$^{-1}$ is significantly exceeded. It should be emphasized, that this observation is only unambiguously accessible by the additional measurements of $\Delta l/l_0$, which allows excluding the frequently discussed influence of thermocouple delays on magnetostructural transitions [30, 31].

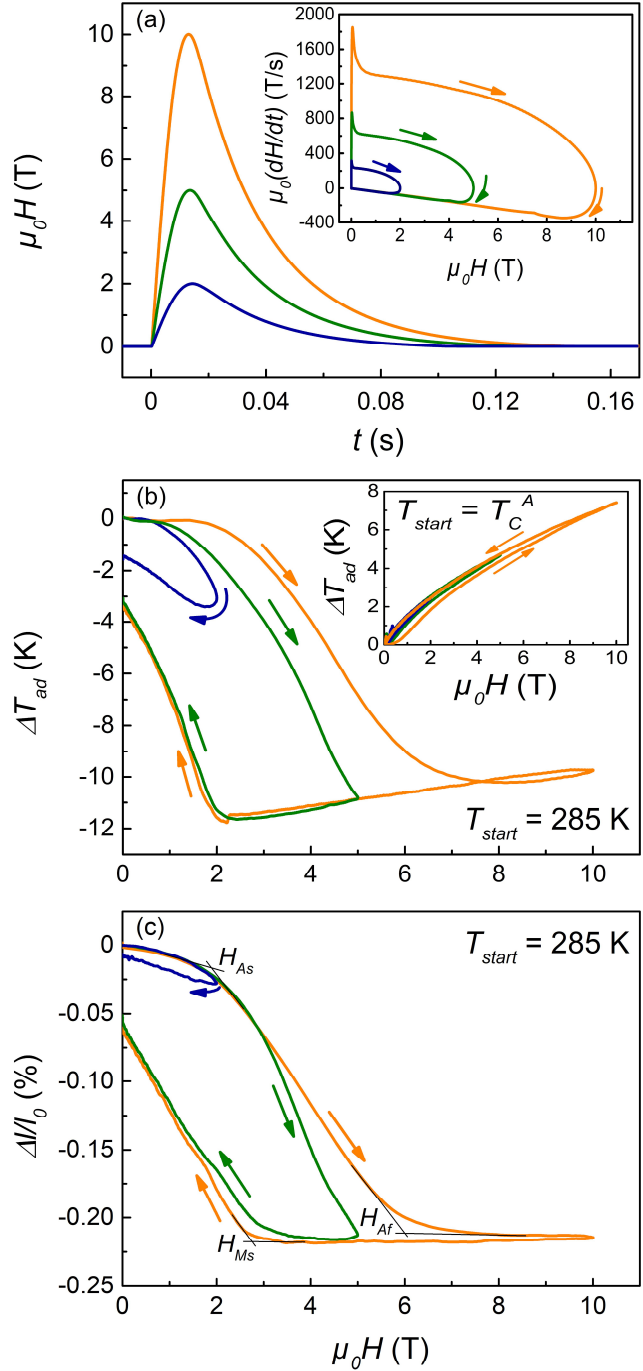

*Figure 6: (a) Magnetic field as a function of time. The inset shows the corresponding magnetic-field sweep rates as a function of the magnetic field. (b) Adiabatic temperature change $\Delta T_{ad}$ and (c) strain $\Delta l/l_0$ depending on magnetic fields of 2 T, 5 T and 10 T. The inset in (b) illustrates the conventional magnetocaloric effect at the Curie-Temperature $T_C^A$ of the austenite. The determination of the critical fields $H_{As}$, $H_{Af}$ and $H_{Ms}$ is exemplarily shown for the 10 T pulse in (c).*

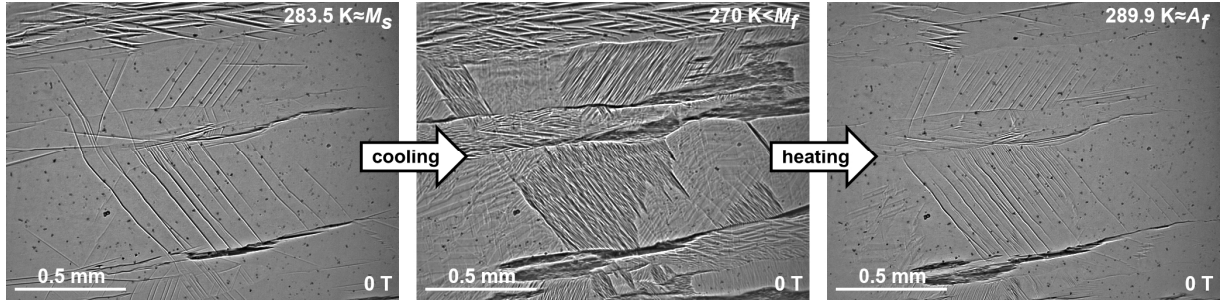

Figure 7: (a) Micrographs of the sample upon cooling at 283.5 K ≈ $M_s$ and 270 K (below $M_f$) and during the subsequent heating at 289.9 K ≈ $A_f$.

## 4. Conclusion

Simultaneous $\Delta T_{ad}$ and $\Delta l/l_0$ measurements were performed in pulsed magnetic fields of 2 T, 5 T and 10 T on a Ni-Mn-In Heusler compound. A strong coupling of both quantities could be observed upon the reverse martensitic transformation resulting in a $\Delta T_{ad}$ of -10 K and a $\Delta l/l_0$ of -0.22 % in 10 T. Upon the field removal, an additional increase of the $|\Delta T_{ad}|$ by 15 % down to -11.5 K occurs due to the conventional MCE of the Curie temperature of the austenite phase $T_C^A$. The rise of $|\Delta T_{ad}|$ can be utilized in the so-called "exploiting-hysteresis cycle" [15] by using materials with a sufficiently large thermal hysteresis, which completely hinders the magnetostructural back transformation.

By different maximum field-sweep rates of 1850 Ts$^{-1}$, 865 Ts$^{-1}$ and 316 Ts$^{-1}$ coming along with the 10 T, 5 T and 2 T pulses, the time dependence of the reverse martensitic transformation could be shown. With the simultaneous measurements of $\Delta T_{ad}$ and $\Delta l/l_0$ it could be demonstrated that the onset of the reverse martensitic transformation is not affected in the studied field range. However, at field rates significantly larger than 865 Ts$^{-1}$ a retardation of the reverse martensitic transformation with respect to the magnetic field was observed which we attribute to the annihilation process of the retained martensite. In consequence, an increasing field hysteresis and higher saturation fields of the metamagnetic transition arise when the critical field-sweeping rate is exceeded.

In conclusion, our findings make clear that the transformation kinetics of Ni-Mn-In and other first-order materials become time-dependent in high magnetic field-sweep rates. This is an important factor to be considered when designing a multicaloric cooling application, e.g. using the "exploiting-hysteresis cycle".

## 5. Acknowledgements

The work was supported by funding from the European Research Council (ERC) under the European Union's Horizon 2020 research and innovation programme (grant no. 743116—project Cool Innov). We thank the HLD at HZDR, member of the European Magnetic Field Laboratory (EMFL) and the Helmholtz Association for funding via the Helmholtz-RSF Joint Research Group (Project No. HRSF-0045) and the Deutsche Forschungsgemeinschaft (DFG, German Research Foundation) – Project-ID 405553726 – TRR 270. A.Y. Karpenkov gratefully acknowledges the Russian Science Foundation (Grant No. 18-42-06201)